\documentclass[12pt]{article}
\usepackage{graphicx}

\begin{document}

\title{\textbf{Formation of superheavy nuclei in cold fusion reactions}}

\author{Zhao-Qing Feng$^{1,2}$, Gen-Ming Jin$^{1}$, Jun-Qing Li$^{1}$, Werner
Scheid$^{3}$}
\date{}
\maketitle

\begin{center}
$^{1}${\small \emph{Institute of Modern Physics, Chinese Academy of
Sciences, Lanzhou 730000, China}}\\[0pt]
$^{2}${\small \emph{Gesellschaft f\"{u}r Schwerionenforschung mbH
(GSI) D-64291 Darmstadt, Germany}}\\[0pt]
$^{3}${\small \emph{Institut f\"{u}r Theoretische Physik der
Universit\"{a}t, 35392 Giessen, Germany}}
\end{center}

\begin{abstract}
Within the concept of the dinuclear system (DNS), a dynamical model
is proposed for describing the formation of superheavy nuclei in
complete fusion reactions by incorporating the coupling of the
relative motion to the nucleon transfer process. The capture of two
heavy colliding nuclei, the formation of the compound nucleus and
the de-excitation process are calculated by using an empirical
coupled channel model, solving a master equation numerically and
applying statistical theory, respectively. Evaporation residue
excitation functions in cold fusion reactions are investigated
systematically and compared with available experimental data.
Maximal production cross sections of superheavy nuclei in cold
fusion reactions with stable neutron-rich projectiles are obtained.
Isotopic trends in the production of the superheavy elements Z=110,
112, 114, 116, 118 and 120 are analyzed systematically. Optimal
combinations and the corresponding excitation energies are proposed.
\end{abstract}


\section{INTRODUCTION}

The synthesis of very heavy (superheavy) nuclei is a very important
subject in nuclear physics motivated with respect to the island of
stability which is predicted theoretically, and has obtained much
experimental progress with fusion-evaporation reactions
$\cite{Ho00,Og07}$. The existence of the superheavy nucleus (SHN)
($Z\geq106$) is due to a strong binding shell effect against the
large Coulomb repulsion. However, the shell effect will be reduced
with increasing excitation energy of the formed compound nucleus.
Combinations with a doubly magic nucleus or nearly magic nucleus are
usually chosen due to the larger reaction Q values. Reactions with
$^{208}Pb$ or $^{209}Bi$ targets are proposed firstly by Yu. Ts.
Oganessian et al. to synthesize SHN $\cite{Og75}$. Six new elements
with Z=107-112 were synthesized in cold fusion reactions for the
first time and investigated at GSI (Darmstadt, Germany) with the
heavy-ion accelerator UNILAC and the separator SHIP
$\cite{Ho00,Mu99}$. Recently, experiments on the synthesis of
element 113 in the $^{70}Zn+^{209}Bi$ reaction have been performed
successfully at RIKEN (Tokyo, Japan) $\cite{Mo04}$. Superheavy
elements Z=113-116, 118 were synthesized at FLNR in Dubna (Russia)
with double magic nucleus $^{48}Ca$ bombarding actinide nuclei
$\cite{Og99}$. Reasonable understanding on the formation of SHN in
massive fusion reactions is still a challenge for theory.

In accordance with the evolution of two heavy colliding nuclei, the
whole process of the compound nucleus formation and decay is usually
divided into three reaction stages, namely the capture process of
the colliding system to overcome Coulomb barrier, the formation of
the compound nucleus to pass over the inner fusion barrier as well
as the de-excitation of the excited compound nucleus against
fission. The transmission in the capture process depends on the
incident energy and relative angular momentum of the colliding
nuclei, and is the same as in the fusion of light and medium mass
systems. The complete fusion of the heavy system after capture in
competition with quasi-fission is very important in the estimation
of the SHN production. At present it is still difficult to make an
accurate description of the fusion dynamics. After the capture and
the subsequent evolution to form the compound nucleus, the thermal
compound nucleus will decay by the emission of light particles and
$\gamma$-rays against fission. The above three stages will affect
the formation of evaporation residues observed in laboratories. The
evolution of the whole process of massive heavy-ion collisions is
very complicated at near barrier energies. Most of theoretical
approaches on the formation of SHN have a similar viewpoint in the
description of the capture and the de-excitation stages, but there
is no consensus on the compound nucleus formation process. There are
mainly two sorts of models, whether the compound nucleus is formed
along the radial variable (internuclear distance) or by nucleon
transfer at the minimum position of the interaction potential after
capture of the colliding system. Several transport models have been
established to understand the fusion mechanism of two heavy
colliding nuclei leading to SHN formation, such as the macroscopic
dynamical model $\cite{Sw80,Bj82}$, the fluctuation-dissipation
model $\cite{Ar99}$, the concept of nucleon collectivization
$\cite{Za01}$ and the dinuclear system model $\cite{Ad97}$. With
these models experimental data can be reproduced and some new
results have been predicted. The models differ from each other, and
sometimes contradictory physical ideas are used.

Further improvements on the mentioned models have to be made. Here
we use an improved dinuclear system model (DNS), in which the
nucleon transfer is coupled with the relative motion and the barrier
distribution of the colliding system is included. We present a new
and extended investigation of the production of superheavy nuclei in
lead-based cold fusion reactions. For that we make use of a
formalism describing the nucleon transfer with a set of
microscopically derived master equations.

In Sec. $2$ we give a description on the DNS model. Calculated
results of fusion dynamics and SHN production in cold fusion
reactions are given in Sec. $3$. In Sec. $4$ conclusions are
discussed.

\section{DINUCLEAR SYSTEM MODEL}

The dinuclear system (DNS) is a molecular configuration of two
touching nuclei which keep their own individuality $\cite{Ad97}$.
Such a system has an evolution along two main degrees of freedom:
(i) the relative motion of the nuclei in the interaction potential
to form the DNS and the decay of the DNS (quasi-fission process)
along the R degree of freedom (internuclear motion), (ii) the
transfer of nucleons in the mass asymmetry coordinate
$\eta=(A_{1}-A_{2})/(A_{1}+A_{2})$ between two nuclei, which is a
diffusion process of the excited systems leading to the compound
nucleus formation. Off-diagonal diffusion in the surface $(A_{1},R)$
is not considered since we assume the DNS is formed at the minimum
position of the interaction potential of two colliding nuclei. In
this concept, the evaporation residue cross section is expressed as
a sum over partial waves with angular momentum $J$ at the
centre-of-mass energy $E_{c.m.}$,
\begin{equation}
\sigma_{ER}(E_{c.m.})=\frac{\pi \hbar^{2}}{2\mu
E_{c.m.}}\sum_{J=0}^{J_{max}}(2J+1)
T(E_{c.m.},J)P_{CN}(E_{c.m.},J)W_{sur}(E_{c.m.},J).
\end{equation}
Here, $T(E_{c.m.},J)$ is the transmission probability of the two
colliding nuclei overcoming the Coulomb potential barrier in the
entrance channel to form the DNS. In the same manner as in the
nucleon collectivization model $\cite{Za01}$, the transmission
probability $T$ is calculated by using the empirical coupled channel
model, which can reproduce very well available experimental capture
cross sections $\cite{Za01,Fe06}$. $P_{CN}$ is the probability that
the system will evolve from a touching configuration into the
compound nucleus in competition with quasi-fission of the DNS and
fission of the heavy fragment. The last term is the survival
probability of the formed compound nucleus, which can be estimated
with the statistical evaporation model by considering the
competition between neutron evaporation and fission $\cite{Fe06}$.
We take the maximal angular momentum as $J_{max}=30$ since the
fission barrier of the heavy nucleus disappears at high spin
$\cite{Re00}$.

In order to describe the fusion dynamics as a diffusion process in
mass asymmetry, the analytical solution of the Fokker-Planck
equation $\cite{Ad97}$ and the numerical solution of the master
equation $\cite{Li03,Di01}$ have been used, which were also used to
treat deep inelastic heavy-ion collisions. Here, the fusion
probability is obtained by solving a master equation numerically in
the potential energy surface of the DNS. The time evolution of the
distribution function $P(A_{1},E_{1},t)$ for fragment 1 with mass
number $A_{1}$ and excitation energy $E_{1}$ is described by the
following master equation $\cite{No75,Fe87}$,
\begin{eqnarray}
\frac{d P(A_{1},E_{1},t)}{dt}=\sum_{A_{1}^{\prime
}}W_{A_{1},A_{1}^{\prime}}(t)\left[
d_{A_{1}}P(A_{1}^{\prime},E_{1}^{\prime},t)-d_{A_{1}^{\prime
}}P(A_{1},E_{1},t)\right]-
\nonumber \\
\left[\Lambda^{qf}(\Theta(t))+\Lambda^{fis}(\Theta(t))
\right]P(A_{1},E_{1},t).
\end{eqnarray}
Here $W_{A_{1},A_{1}^{\prime}}$ is the mean transition probability
from the channel $(A_{1},E_{1})$ to
$(A_{1}^{\prime},E_{1}^{\prime})$, while $d_{A_{1}}$ denotes the
microscopic dimension corresponding to the macroscopic state
$(A_{1},E_{1})$. The sum is taken over all possible mass numbers
that fragment $A_{1}^{\prime}$ may take (from $0$ to
$A=A_{1}+A_{2}$), but only one nucleon transfer is considered in the
model with $A_{1}^{\prime }=A_{1}\pm 1$. The excitation energy
$E_{1}$ is the local excitation energy $\varepsilon^{\ast}_{1}$ with
respect to fragment $A_{1}$, which is determined by the dissipation
energy from the relative motion and the potential energy of the
corresponding DNS and will be shown later in Eqs.(8-9). The
dissipation energy is described by the parametrization method of the
classical deflection function $\cite{Wo78,Li81}$. The motion of
nucleons in the interacting potential is governed by the
single-particle Hamiltonian $\cite{Fe06,Li03}$:
\begin{equation}
H(t)=H_{0}(t)+V(t)
\end{equation}
with
\begin{eqnarray}
H_{0}(t)&=&\sum_{K}\sum_{\nu_{K}}\varepsilon_{\nu_{K}}(t)a_{\nu_{K}}^{\dag}(t)a_{\nu_{K}}(t),
\nonumber \\
V(t)&=&\sum_{K,K^{\prime}}\sum_{\alpha_{K},\beta_{K^{\prime}}}u_{\alpha_{K},\beta_{K^{\prime}}}(t)
a_{\alpha_{K}}^{\dag}(t)a_{\beta_{K^{\prime}}}(t)=\sum_{K,K^{\prime}}V_{K,K^{\prime}}(t).
\end{eqnarray}
Here the indices $K, K^{\prime}$ $(K,K^{\prime}=1,2)$ denote the
fragments $1$ and $2$. The quantities $\varepsilon_{\nu_{K}}$ and
$u_{\alpha_{K},\beta_{K^{\prime}}}$ represent the single particle
energies and the interaction matrix elements, respectively. The
single particle states are defined with respect to the centers of
the interacting nuclei and are assumed to be orthogonalized in the
overlap region. So the annihilation and creation operators are
dependent on time. The single particle matrix elements are
parameterized by
\begin{equation}
u_{\alpha_{K},\beta_{K^{\prime}}}(t)=U_{K,K^{\prime}}(t)\left\{\exp\left[-\frac{1}{2}
\left(\frac{\varepsilon_{\alpha_{K}}(t)-\varepsilon_{\beta_{K^{\prime}}}(t)}
{\Delta_{K,K^{\prime}}(t)}\right)^{2}\right]-
\delta_{\alpha_{K},\beta_{K^{\prime}}}\right\},
\end{equation}
which contains some parameters $U_{K,K^{\prime}}(t)$ and
$\Delta_{K,K^{\prime}}(t)$. The detailed calculation of these
parameters and the mean transition probabilities were described in
Refs. $\cite{Li03,Fe06}$.

The evolution of the DNS along the variable R leads to the
quasi-fission of the DNS. The quasi-fission rate $\Lambda^{qf}$ can
be estimated with the one dimensional Kramers formula
$\cite{Ad03,Gr83}$:
\begin{equation}
\Lambda^{qf}(\Theta(t))=\frac{\omega}{2\pi\omega^{B_{qf}}}\left(\sqrt{\left(\frac{\Gamma}
{2\hbar}\right)^{2}+(\omega^{B_{qf}})^{2}}-\frac{\Gamma}
{2\hbar}\right)\exp\left(-\frac{B_{qf}(A_{1},A_{2})}{\Theta(t)}\right).
\end{equation}
Here the quasi-fission barrier measures the depth of the pocket of
the interaction potential. The local temperature is given by the
Fermi-gas expression $\Theta=\sqrt{\varepsilon^{\star}/a}$
corresponding to the local excitation energy $\varepsilon^{\star}$
and level density parameter $a=A/12$ $MeV^{-1}$. $\omega^{B_{qf}}$
is the frequency of the inverted harmonic oscillator approximating
the interaction potential of two nuclei in R around the top of the
quasi-fission barrier, and $\omega$ is the frequency of the harmonic
oscillator approximating the potential in R at the bottom of the
pocket. The quantity $\Gamma$ denotes the double average width of
the contributing single-particle states, which determines the
friction coefficients:
$\gamma_{ii^{\prime}}=\frac{\Gamma}{\hbar}\mu_{ii^{\prime}}$, with
$\mu_{ii^{\prime}}$ being the inertia tensor. Here we use constant
values $\Gamma=2.8$ MeV, $\hbar\omega^{B_{qf}}=2.0$ MeV and
$\hbar\omega=3.0$ MeV for the following reactions. The Kramers
formula is derived at the quasi-stationary condition of the
temperature $\Theta(t)<B_{qf}(A_{1},A_{2})$. However, the numerical
calculation in Ref. $\cite{Gr83}$ indicated that Eq.(6) is also
available at the condition of $\Theta(t)>B_{qf}(A_{1},A_{2})$. In
the reactions of synthesizing SHN, there is the possibility of the
fission of the heavy fragment in the DNS. Since the fissility
increases with the charge number of the nucleus, the fission of the
heavy fragment can affect the quasi-fission and fusion when the DNS
evolves towards larger mass asymmetry. The fission rate
$\Lambda^{fis}$ can also be treated with the one-dimensional Kramers
formula $\cite{Ad03}$
\begin{equation}
\Lambda^{fis}(\Theta(t))=\frac{\omega_{g.s.}}{2\pi\omega_{f}}\left(\sqrt{\left(\frac{\Gamma_{0}}
{2\hbar}\right)^{2}+\omega_{f}^{2}}-\frac{\Gamma_{0}}
{2\hbar}\right)\exp\left(-\frac{B_{f}(A_{1},A_{2})}{\Theta(t)}\right),
\end{equation}
where $\omega_{g.s.}$ and $\omega_{f}$ are the frequencies of the
oscillators approximating the fission-path potential at the ground
state and on the top of the fission barrier for nucleus $A_{1}$ or
$A_{2}$ (larger fragment), respectively. Here, we take
$\hbar\omega_{g.s.}=\hbar\omega_{f}=1.0$ MeV, $\Gamma_{0}=2$ MeV.
The fission barrier is calculated as a sum of a macroscopic part and
the shell correction used in Refs. $\cite{Ad00}$. The fission of the
heavy fragment is not in favor of the diffusion of the system to
light fragment distribution. Therefore, it leads to a slightly
decrease of the fusion probability (seeing Eq.(17)).

In the relaxation process of the relative motion, the DNS will be
excited due to the dissipation of the relative kinetic energy. The
excited system opens a valence space $\Delta\varepsilon_{K}$ in
fragment $K (K=1,2)$, which has a symmetrical distribution around
the Fermi surface. Only the particles in the states within the
valence space are actively involved in excitation and transfer. The
averages on these quantities are performed in the valence space:
\begin{equation}
\Delta\varepsilon_{K}=\sqrt{\frac{4\varepsilon^{\ast}_{K}}{g_{K}}},
\varepsilon^{\ast}_{K}=\varepsilon^{\ast}\frac{A_{K}}{A},
g_{K}=\frac{A_{K}}{12}
\end{equation}
where $\varepsilon^{\ast}$ is the local excitation energy of the
DNS, which provides the excitation energy for the mean transition
probability. There are $N_{K}=g_{K}\Delta\varepsilon_{K}$ valence
states and $m_{K}=N_{K}/2$ valence nucleons in the valence space
$\Delta\varepsilon_{K}$, which give the dimension
$d(m_{1},m_{2})=\left(\begin{array}{c}
N_{1} \\
m_{1}
\end{array}\right)
\left(\begin{array}{c}
N_{2} \\
m_{2}
\end{array}\right)$. The local
excitation energy is defined as
\begin{equation}
\varepsilon^{\ast}=E_{x}-\left(U(A_{1},A_{2})-U(A_{P},A_{T})\right).
\end{equation}
Here $U(A_{1},A_{2})$ and $U(A_{P},A_{T})$ are the driving
potentials of fragments $A_{1},A_{2}$ and fragments $A_{P},A_{T}$
(at the entrance point of the DNS), respectively. The excitation
energy $E_{x}$ of the composite system is converted from the
relative kinetic energy loss, which is related to the Coulomb
barrier $B$ $\cite{Fe07}$ and determined for each initial relative
angular momentum $J$ by the parametrization method of the classical
deflection function $\cite{Wo78,Li81}$. So $E_{x}$ is coupled with
the relative angular momentum.

The potential energy surface (PES, i.e. the driving potential) of
the DNS is given by
\begin{equation}
U(A_{1},A_{2},J,\textbf{R};\beta_{1},\beta_{2},\theta_{1},\theta_{2})=B(A_{1})+B(A_{2})-
\left[B(A)+V^{CN}_{rot}(J)\right]+V(A_{1},A_{2},J,\textbf{R};\beta_{1},\beta_{2},\theta_{1},\theta_{2})
\end{equation}
with $A_{1}+A_{2}=A$. Here $B(A_{i}) (i=1,2)$ and $B(A)$ are the
negative binding energies of the fragment $A_{i}$ and the compound
nucleus $A$, respectively, in which the shell and the pairing
corrections are included reasonably. $V^{CN}_{rot}$ is the rotation
energy of the compound nucleus. $\beta_{i}$ represent quadrupole
deformations of the two fragments. $\theta_{i}$ denote the angles
between the collision orientations and the symmetry axes of deformed
nuclei. The interaction potential between fragment $1(Z_{1},A_{1})$
and $2(Z_{2},A_{2})$ includes the nuclear, Coulomb and centrifugal
parts as
\begin{equation}
V(A_{1},A_{2},J,\textbf{R};\beta_{1},\beta_{2},\theta_{1},\theta_{2})=V_{N}(A_{1},A_{2},\textbf{R};\beta_{1},\beta_{2},\theta_{1},\theta_{2})+
V_{C}(A_{1},A_{2},\textbf{R};\beta_{1},\beta_{2},\theta_{1},\theta_{2})+\frac{J(J+1)\hbar^{2}}{2\mu\textbf{R}^{2}},
\end{equation}
where the reduced mass is given by $\mu=m\cdot A_{1}A_{2}/A$ with
the nucleon mass $m$. The nuclear potential is calculated using the
double-folding method based on Skyrme interaction force without
considering the momentum and the spin dependence as $\cite{Ad96}$
\begin{eqnarray}
V_{N}=C_{0}\left\{\frac{F_{in}-F_{ex}}{\rho_{0}}\left[\int\rho_{1}^{2}(\textbf{r})\rho_{2}(\textbf{r}-\textbf{R})d\textbf{r}+
\int\rho_{1}(\textbf{r})\rho_{2}^{2}(\textbf{r}-\textbf{R})d\textbf{r}\right]+F_{ex}\int\rho_{1}(\textbf{r})\rho_{2}(\textbf{r}-\textbf{R})d\textbf{r}\right\},
\end{eqnarray}
with
\begin{equation}
F_{in,ex}=f_{in,ex}+f_{in,ex}^{\prime}\frac{N_{1}-Z_{1}}{A_{1}}\frac{N_{2}-Z_{2}}{A_{2}},
\end{equation}
which is dependent on the nuclear densities and on the orientations
of deformed nuclei in the collision $\cite{Li05}$. The parameters
$C_{0}=300 MeV\cdot fm^{3}$, $f_{in}=0.09$, $f_{ex}=-2.59$,
$f_{in}^{\prime}=0.42$, $f_{ex}^{\prime}=0.54, \rho_{0}=0.16
fm^{-3}$ are used in the calculation. The Woods-Saxon density
distributions are expressed for two nuclei as
\begin{equation}
\rho_{1}(\textbf{r})=\frac{\rho_{0}}{1+\exp[(\textbf{r}-\Re_{1}(\theta_{1}))/a_{1}]},
\end{equation}
and
\begin{equation}
\rho_{2}(\textbf{r}-\textbf{R})=\frac{\rho_{0}}{1+\exp[(|\textbf{r}-\textbf{R}|-\Re_{2}(\theta_{2}))/a_{2}]}.
\end{equation}
Here $\Re_{i}(\theta_{i})$ $(i=1,2)$ are the surface radii of the
nuclei with
$\Re_{i}(\theta_{i})=R_{i}(1+\beta_{i}Y_{20}(\theta_{i}))$, and the
spheroidal radii $R_{i}$. The parameters $a_{i}$ represent the
surface diffusion coefficients, which are taken 0.55 fm in the
calculation. The Coulomb potential is obtained by Wong's formula
$\cite{Wo73}$, which agrees well with the double-folding procedure.
In the actual calculation, the distance $\textbf{R}$ between the
centers of the two fragments is chosen to be the value which gives
the minimum of the interaction potential, in which the DNS is
considered to be formed. So the PES depends only on the mass
asymmetry degree of freedom $\eta$, which gives the driving
potential of the DNS as shown in Fig.1 for the reaction
$^{70}Zn+^{208}Pb$ at the tip-tip, the belly-belly and at the fixed
$(0^{o},0^{o})$ and $(90^{o},90^{o})$ orientations. Here, we should
note that the tip-tip orientation is different with $(0^{o},0^{o})$.
We rotate $\frac{\pi}{2}$ for the fragment with negative quadrupole
deformation. However, the orientation angle $\theta_{i}$ is fixed
for all fragments. The same procedure is taken for the belly-belly
and $(90^{o},90^{o})$. The Businaro-Gallone (B.G.) point marks the
maximum position of the driving potential on the left side of the
initial combination $\eta_{i}$. Some averaging over all orientations
should be carried out in the nucleon transfer process. However, the
tip-tip orientation which gives the minimum of the PES is in favor
of nucleon transfer and is chosen in the calculation. For the
reaction $^{70}Zn+^{208}Pb$, the tip-tip orientation
($B_{fus}$=20.98 MeV) has lower inner fusion barrier than the
belly-belly orientation ($B_{fus}$=25.71 MeV). However, the
belly-belly orientation appears an obvious hump towards symmetric
combinations (reducing $|\eta_{i}|$), which is in favor of the
compound nucleus formation against the quasi-fission. Both of the
two factors may affect the values of $P_{CN}$ (seeing Eq.(17)). In
Fig.2 we show the comparison of the formation probability of the
compound nucleus in the reaction $^{70}Zn+^{208}Pb$ as functions of
angular momenta ($E_{c.m.}$=254.08 MeV, $E^{\ast}_{CN}$=12 MeV) and
incident c.m. energies (J=0) at the tip-tip and the belly-belly
orientations, respectively. The effects of the collision
orientations on the fusion cross section were also studied in detail
by A. Nasirov et al. $\cite{Na05}$ for deformed combination systems.

%%%%%%%%%%%%%%%%%%%%%%%%%%%%%%%%%%%%%%% figure 1 %%%%%%%%%%%%%%%%%%%%%%%%
\begin{figure}
\begin{center}
{\includegraphics*[width=0.8\textwidth]{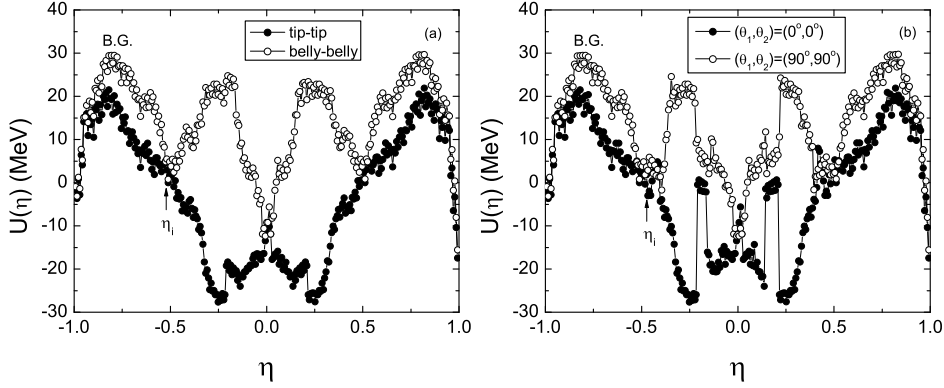}}
\end{center}
\caption{The driving potential of the DNS for the reaction
$^{70}Zn+^{208}Pb$ as a function of the mass asymmetry $\eta$ at the
different orientations.}
\end{figure}
%%%%%%%%%%%%%%%%%%%%%%%%%%%%%%%%%%%%%%%%%%%%%%%%%%%%%%%%%%%%%%%%%%%%%%%%%

%%%%%%%%%%%%%%%%%%%%%%%%%%%%%%%%%%%%%%% figure 2 %%%%%%%%%%%%%%%%%%%%%%%%
\begin{figure}
\begin{center}
{\includegraphics*[width=0.8\textwidth]{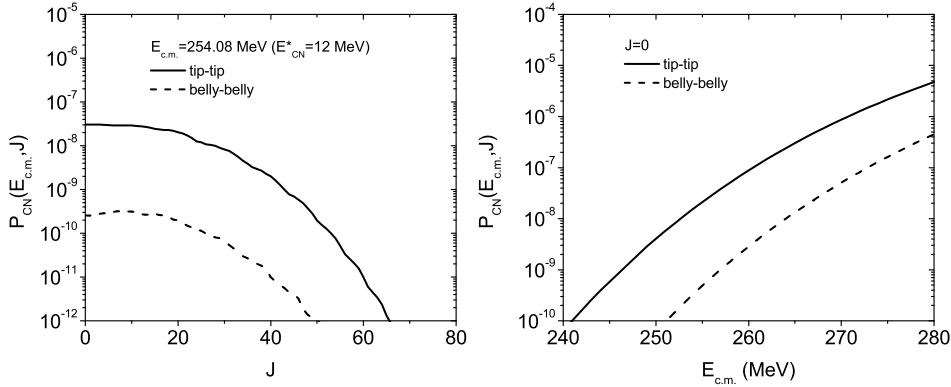}}
\end{center}
\caption{Dependence of the fusion probability on angular momenta and
incident c.m. energies in the reaction $^{70}Zn+^{208}Pb$ at the
tip-tip and the belly-belly orientations, respectively.}
\end{figure}
%%%%%%%%%%%%%%%%%%%%%%%%%%%%%%%%%%%%%%%%%%%%%%%%%%%%%%%%%%%%%%%%%%%%%%%%%

After reaching the time of reaction in the evolution of
$P(A_{1},E_{1},t)$, all those components on the left side of the
B.G. point as shown in Fig.1 (a) contribute to the compound nucleus
formation. The hindrance in the diffusion process by nucleon
transfer to form the compound nucleus is the inner fusion barrier
$B_{fus}$, which is defined as the difference of the driving
potential at the B.G. point and at the entrance position. Nucleon
transfer to more symmetric fragments will be in favor of
quasi-fission. The formation probability of the compound nucleus at
Coulomb barrier $B$ (here the barrier distribution $f(B)$ is
considered) and angular momentum $J$ is given by
\begin{equation}
P_{CN}(E_{c.m.},J,B)=\sum_{A_{1}=1}^{A_{BG}}P(A_{1},E_{1},\tau
_{int}(E_{c.m.},J,B)).
\end{equation}
Here the interaction time $\tau _{int}(E_{c.m.},J,B)$ is obtained
using the deflection function method $\cite{Li83}$. We obtain the
fusion probability as
\begin{equation}
P_{CN}(E_{c.m.},J)=\int f(B)P_{CN}(E_{c.m.},J,B)dB,
\end{equation}
where the barrier distribution function is taken in asymmetric
Gaussian form $\cite{Za01,Fe06}$. So the fusion cross section is
written as
\begin{equation}
\sigma_{fus}(E_{c.m.})=\frac{\pi \hbar^{2}}{2\mu
E_{c.m.}}\sum_{J=0}^{\infty}(2J+1) T(E_{c.m.},J)P_{CN}(E_{c.m.},J).
\end{equation}

The survival probability of the excited compound nucleus in the
cooling process by means of the neutron evaporation in competition
with fission is expressed as following:
\begin{equation}
W_{sur}(E_{CN}^{\ast},x,J)=P(E_{CN}^{\ast},x,J)\prod\limits_{i=1}^{x}\left(
\frac{\Gamma _{n}(E_{i}^{\ast},J)}{\Gamma
_{n}(E_{i}^{\ast},J)+\Gamma _{f}(E_{i}^{\ast},J)}\right) _{i},
\end{equation}
where the $E_{CN}^{\ast}, J$ are the excitation energy and the spin
of the compound nucleus, respectively. $E_{i}^{\ast}$ is the
excitation energy before evaporating the $i$th neutron, which has
the relation:
\begin{equation}
E_{i+1}^{\ast}=E_{i}^{\ast}-B_{i}^{n}-2T_{i},
\end{equation}
with the initial condition $E_{1}^{\ast}=E_{CN}^{\ast}$. $B_{i}^{n}$
is the separation energy of the $i$th neutron. The nuclear
temperature $T_{i}$ is given by $E_{i}^{\ast}=aT_{i}^{2}-T_{i}$ with
the level density parameter $a$. $P(E_{CN}^{\ast},x,J)$ is the
realization probability of emitting $x$ neutrons. The widths of
neutron evaporation and fission are calculated using the statistical
model. The details can be found in Ref. $\cite{Fe06}$.

\section{RESULTS AND DISCUSSION}
\subsection{Evaporation residue cross sections}

The evaporation residues observed in laboratories by the consecutive
$\alpha$ decay are mainly produced by the complete fusion reactions,
in which the fusion dynamics and the structure properties of the
compound nucleus affects their production. Within the framework of
the DNS model, we calculated the evaporation residue cross sections
producing SHN Z=110-113 in cold fusion reactions as shown in Fig.3,
and compared them with GSI data for 110-112 $\cite{Ho00}$ and RIKEN
results $\cite{Mo04}$ for 113. The excitation energy is obtained by
$E^{\ast}_{CN}=E_{c.m.}+Q$, where $E_{c.m.}$ is the incident energy
in the center-of-mass system. The $Q$ value is given by $Q=\Delta
M_{P}+\Delta M_{T}-\Delta M_{C}$, and the corresponding mass defects
are taken from Ref.$\cite{Mo95}$ for projectile, target and compound
nucleus, respectively. Usually, neutron-rich projectiles are used to
synthesize SHN experimentally, such as $^{64}Ni$ and $^{70}Zn$,
which can enhance the survival probability $W_{sur}$ in Eq.(1) of
the formed compound nucleus due to smaller neutron separation
energy. The maximal production cross sections from Ds to 113 are
reduced rapidly because the inner fusion barrier is increasing.
Within error bars the experimental results can be reproduced very
well. There is no other adjustable parameters in the calculation.
Within the same scheme, we analyzed the evaporation residue
excitation functions with projectiles $^{73}Ge$, $^{82}Se$,
$^{86}Kr$ and $^{88}Sr$ to produce superheavy elements Z=114, 116,
118, 120 in Fig.4. An upper-limit for the cross section producing
118 was obtained in Berkeley $\cite{Gr03}$.

%%%%%%%%%%%%%%%%%%%%%%%%%%%%%%%%%%%%%%% figure 3 %%%%%%%%%%%%%%%%%%%%%%%%
\begin{figure}
\begin{center}
{\includegraphics*[width=0.8\textwidth]{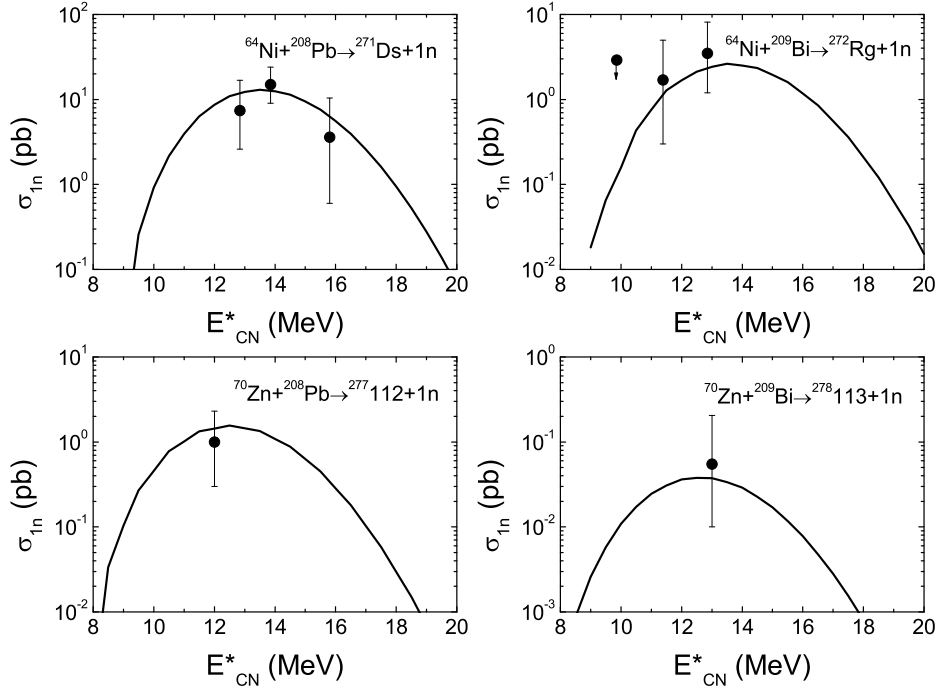}}
\end{center}
\caption{Comparison of the calculated evaporation residue excitation
functions and the experimental data to synthesize superheavy
elements Z=110-113 in cold fusion reactions.}
\end{figure}
%%%%%%%%%%%%%%%%%%%%%%%%%%%%%%%%%%%%%%%%%%%%%%%%%%%%%%%%%%%%%%%%%%%%%%%%%

%%%%%%%%%%%%%%%%%%%%%%%%%%%%%%%%%%%%% figure 4 %%%%%%%%%%%%%%%%%%%%%%%%%%
\begin{figure}
\begin{center}
{\includegraphics*[width=0.8\textwidth]{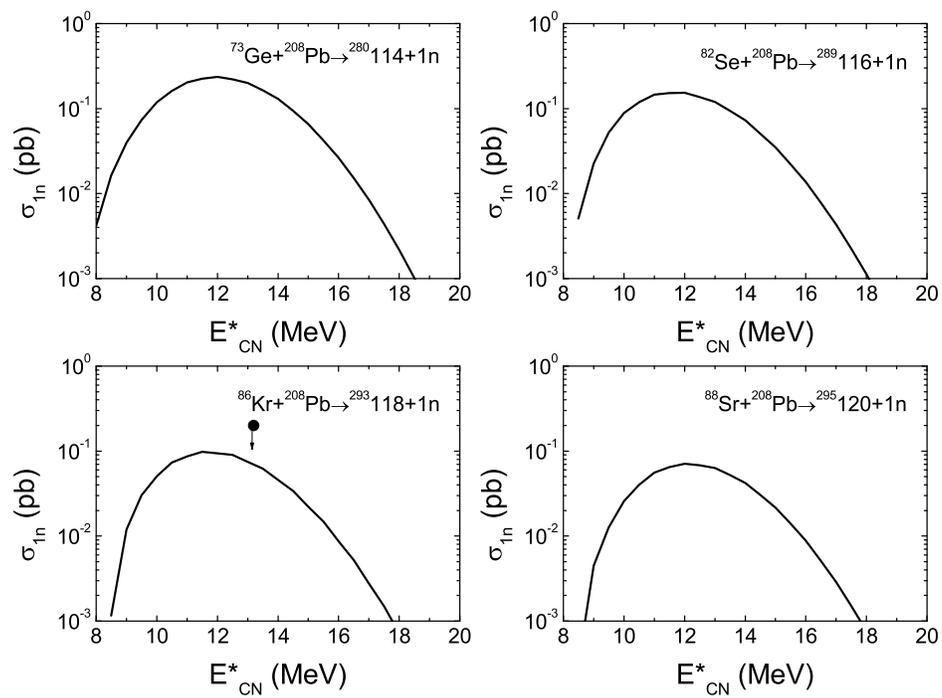}}
\end{center}
\caption{The same as in Fig.3, but for projectiles $^{73}Ge$,
$^{82}Se$, $^{86}Kr$ and $^{88}Sr$ in cold fusion reactions to
produce superheavy elements Z=114, 116, 118, 120.}
\end{figure}
%%%%%%%%%%%%%%%%%%%%%%%%%%%%%%%%%%%%%%%%%%%%%%%%%%%%%%%%%%%%%%%%%%%%%%%%%

In Fig.5 we show the comparison of the calculated maximal production
cross sections of superheavy elements Z=102-120 in cold fusion
reactions by evaporating one neutron with experimental data
$\cite{Ho00,Mu99}$. The production cross sections decrease rapidly
with increasing the charge number of the synthesized compound
nucleus, such as from 0.2 $\mu b$ for the reaction
$^{48}Ca+^{208}Pb$ to 1 pb for $^{70}Zn+^{208}Pb$, and around even
below 0.1 pb for synthesizing Z$\geq$113. It seems to be difficult
to synthesize superheavy elements Z$\geq$113 in cold fusion
reactions at the present facilities. The calculated results are in
good agreement with the experimental data. In the DNS concept, the
inner fusion barrier increases with reducing mass asymmetry, which
leads to a decrease of the formation probability of the compound
nucleus as shown in Fig.6. On the other hand, the quasi-fission and
the fission of the heavy fragments in the nuclear transfer process
become more and more important if the mass asymmetry ($|\eta_{i}|$)
of the projectile-target combination is decreasing, which also
reduce the formation probability. There appears a little increase
for Z$\geq$118, which is related to the decreased inner fusion
barriers of the three systems. The survival of the thermal compound
nucleus in the fusion reactions are mainly affected by the neutron
evaporation energy, the fission barrier and the level density. The
survival probability has strong structure effects as shown in Fig.6.
Accurate calculation of the survival probability is very necessary
to obtain reasonable evaporation residue cross sections.

%%%%%%%%%%%%%%%%%%%%%%%%%%%%%%%%%%%% figure 5 %%%%%%%%%%%%%%%%%%%%%%%%%%%
\begin{figure}
\begin{center}
{\includegraphics*[width=0.8\textwidth]{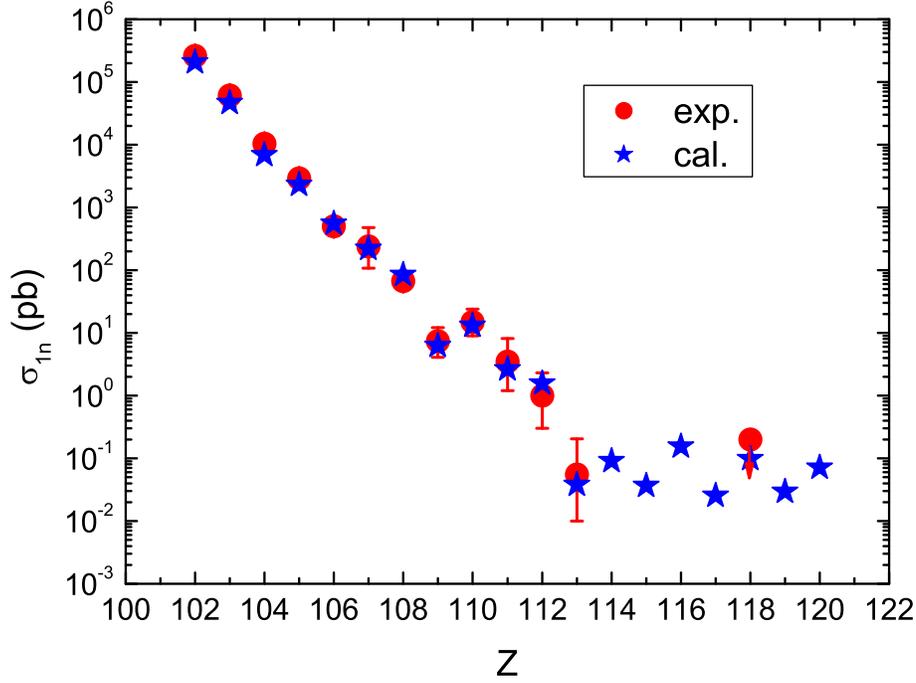}}
\end{center}
\caption{(Color online) Maximal production cross sections of
superheavy elements Z=102-120 in cold fusion reactions based
$^{208}Pb$ and $^{209}Bi$ targets with projectile nuclei $^{48}Ca$,
$^{50}Ti$, $^{54}Cr$, $^{58}Fe$, $^{64}Ni$, $^{70}Zn$, $^{76}Ge$,
$^{82}Se$, $^{86}Kr$ and $^{88}Sr$, and compared with experimental
data.}
\end{figure}
%%%%%%%%%%%%%%%%%%%%%%%%%%%%%%%%%%%%%%%%%%%%%%%%%%%%%%%%%%%%%%%%%%%%%%%%%

%%%%%%%%%%%%%%%%%%%%%%%%%%%%%%%%%%%% figure 6 %%%%%%%%%%%%%%%%%%%%%%%%%%%
\begin{figure}
\begin{center}
{\includegraphics*[width=0.8\textwidth]{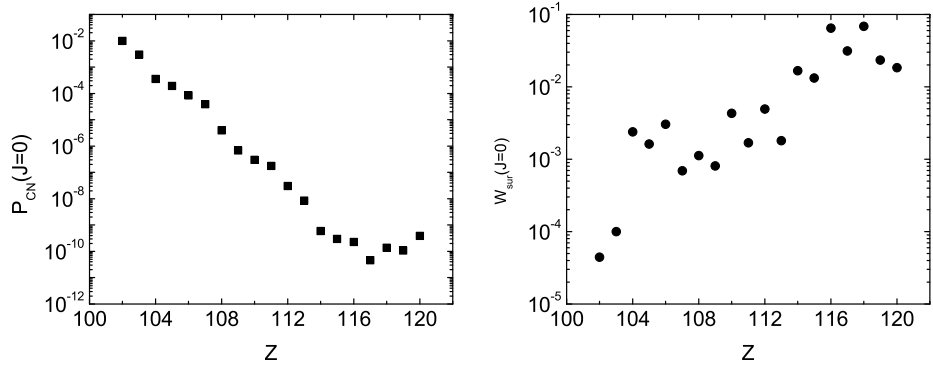}}
\end{center}
\caption{The fusion and the survival probabilities at J=0 as
functions of the charge numbers of the compound nuclei with the same
combinations as stated in the caption of Fig.5.}
\end{figure}
%%%%%%%%%%%%%%%%%%%%%%%%%%%%%%%%%%%%%%%%%%%%%%%%%%%%%%%%%%%%%%%%%%%%%%%%%

\subsection{Isotopic dependence of the production cross sections}

The production of the SHN depends on the isotopic combination of the
target and projectile in the cold fusion reactions. For example, the
maximal cross section is $3.5\pm^{2.7}_{1.8}$ pb for the reaction
$^{62}Ni+^{208}Pb\rightarrow ^{269}Ds+1n$, however $15\pm^{9}_{6}$
pb for the reaction $^{64}Ni+^{208}Pb\rightarrow ^{271}Ds+1n$
$\cite{Ho00,Ho95}$. Further investigations on the isotopic trends
are very necessary for predicting the optimal combinations and the
optimal excitation energies (incident energies) to synthesize SHN.
In Fig.7 we show the calculated isotopic trends in producing
superheavy elements Z=110, 112 for the reactions $^{A}Ni+^{208}Pb$
and $^{A}Zn+^{208}Pb$ (squares with lines), and compare them with
the results of G.G. Adamian et al. $\cite{Ad04}$ (diamonds and
triangles) and the available experimental data $\cite{Ho00}$
(circles with error bars). We find that the isotopes $^{63,64,65}Ni$
and $^{67,70}Zn$ are suitable to synthesize superheavy elements 110
and 112, respectively. The isotopes $^{64}Ni$ and $^{67}Zn$ have
larger production cross section, which is consistent with the
results of G.G. Adamian et al. But for other isotopes, the two
methods give slightly different results. For example, our model
gives that $^{70}Zn$ has lager cross section to produce elements 110
than the isotope $^{68}Zn$. However, the opposite trend is obtained
by G.G. Adamian et al. Therefore, it need more accurate description
on the three stages of the formation of SHN. Further experimental
data is also required to examine the theoretical models. In the DNS
model, the isotopic trends are mainly determined by both the fusion
and survival probabilities. Of course, the transmission probability
of two colliding nuclei can also affect the trends since the initial
quadrupole deformations depend on the isotopes. When the neutron
number of the projectile is increasing, the DNS gets more
symmetrical and the fusion probability decreases if the DNS does not
consist of more stable nuclei due to a higher inner fusion barrier.
A smaller neutron separation energy and a larger shell correction
lead to a larger survival probability. The compound nucleus with
closed neutron shells has larger shell correction energy and neutron
separation energy. With the same procedure, we analyzed the
dependence of the production cross sections on the isotopes Ge and
Se to produce the superheavy elements Z=114, 116 shown in Fig.8 as
well as on the isotopes Kr and Sr to synthesize the superheavy
elements Z=114, 116 with a $^{208}Pb$ target as shown in Fig.9. It
results that the projectiles $^{73}Ge$, $^{79}Se$, $^{85}Kr$ and
$^{87,88}Sr$ are favorable to synthesize the new superheavy elements
Z=114, 116, 118 and 120. The corresponding excitation energies are
also given in the figures. The compound nuclei with neutron-rich
isotopes $^{76}Ge$, $^{80,82}Se$ and $^{84,86}Kr$ are near the
sub-closure at N=172. These compound nuclei have larger one-neutron
separation energies, and the initial combinations smaller mass
asymmetries leading to smaller evaporation residue cross sections.

%%%%%%%%%%%%%%%%%%%%%%%%%%%%%%%%%% figure 7 %%%%%%%%%%%%%%%%%%%%%%%%%%%%%
\begin{figure}
\begin{center}
{\includegraphics*[width=0.8\textwidth]{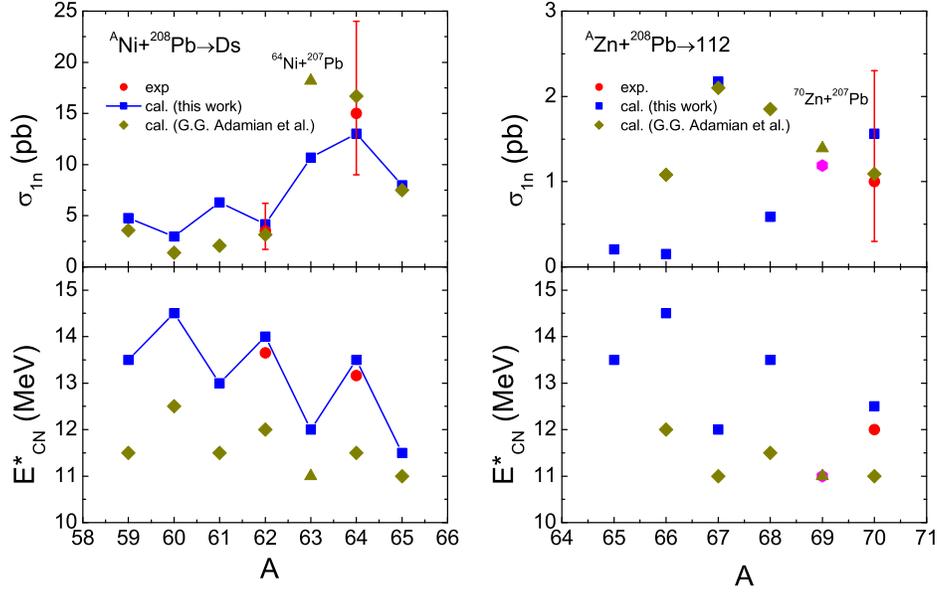}}
\end{center}
\caption{(Color online) Isotopic dependence of the calculated
maximal production cross sections and the corresponding excitation
energies in the synthesis of superheavy elements Z=110, 112 for the
reactions $^{A}Ni+^{208}Pb$ and $^{A}Zn+^{208}Pb$, and compared with
the results of G.G. Adamian et al. $\cite{Ad04}$ and the
experimental data $\cite{Ho00,Mu99}$.}
\end{figure}
%%%%%%%%%%%%%%%%%%%%%%%%%%%%%%%%%%%%%%%%%%%%%%%%%%%%%%%%%%%%%%%%%%%%%%%%%

%%%%%%%%%%%%%%%%%%%%%%%%%%%%%%%%%% figure 8 %%%%%%%%%%%%%%%%%%%%%%%%%%%%%
\begin{figure}
\begin{center}
{\includegraphics*[width=0.8\textwidth]{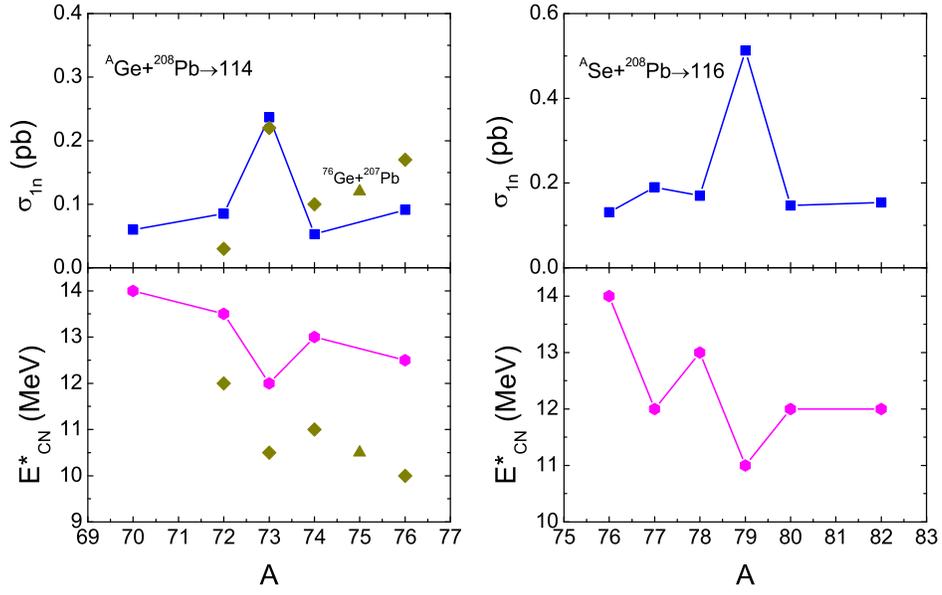}}
\end{center}
\caption{(Color online) The same as in Fig. 7, but for isotopes Ge
and Se to produce superheavy elements Z=114, 116.}
\end{figure}
%%%%%%%%%%%%%%%%%%%%%%%%%%%%%%%%%%%%%%%%%%%%%%%%%%%%%%%%%%%%%%%%%%%%%%%%%

%%%%%%%%%%%%%%%%%%%%%%%%%%%%%%%%%% figure 9 %%%%%%%%%%%%%%%%%%%%%%%%%%%%%
\begin{figure}
\begin{center}
{\includegraphics*[width=0.8\textwidth]{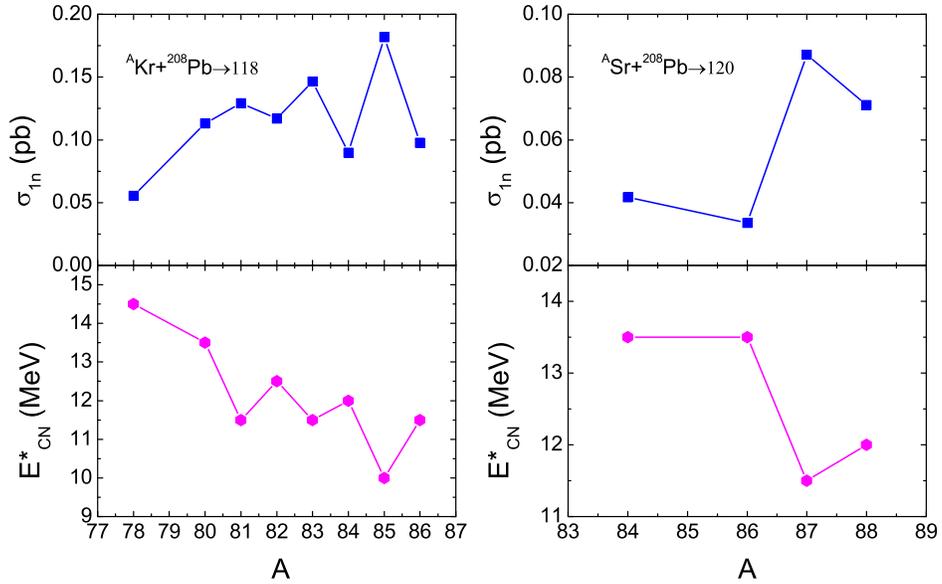}}
\end{center}
\caption{(Color online) The same as in Fig. 7, but for isotopes Kr
and Sr based $^{208}Pb$ target.}
\end{figure}
%%%%%%%%%%%%%%%%%%%%%%%%%%%%%%%%%%%%%%%%%%%%%%%%%%%%%%%%%%%%%%%%%%%%%%%%%

\section{CONCLUSIONS}

Within the DNS concept, a dynamical model is worked out for
describing the production of superheavy residues in the
fusion-evaporation reactions, in which the formation of the
superheavy compound nucleus is described by a master equation which
is solved numerically and includes the quasi-fission of the DNS and
the fission of the heavy fragments in the nucleon transfer process.
By using the DNS model, the fusion dynamics and the evaporation
residue excitation functions in cold fusion reactions are
investigated systematically. The calculated results are in good
agreement with available experimental data within error bars.
Isotopic trends in the production of superheavy elements are
analyzed systematically. It is shown that the isotopes
$^{63,64,65}Ni$, $^{67,70}Zn$, $^{73}Ge$, $^{79}Se$, $^{85}Kr$ and
$^{87,88}Sr$ are favorable to produce the superheavy elements Z=110,
112, 114, 116, 118 and 120 at the stated excitation energies.

The physical nature of the synthesis of heavy fissile nuclei in
massive fusion reactions is very complicated, which involves not
only certain quantities which crucially influence the whole process,
but also the dynamics of the process is important. The coupling of
the dynamic deformation and the nucleon transfer in the course of
overcoming the multi-dimensional PES has to be considered in the DNS
model. The height of the fission barrier for the heavy or superheavy
nuclei should be more studied, which is mainly determined by the
shell correction energies at the ground state and at the saddle
point $\cite{Sm95}$. It plays an very important role in the
calculation of the survival probability. Further work is in
progress.

\section{ACKNOWLEDGMENTS}

One of authors (Z.-Q. Feng) is grateful to Prof. H. Feldmeier, Dr.
G.G. Adamian and Dr. N.V. Antonenko for fruitful discussions and
help. This work was supported by the National Natural Science
Foundation of China under Grant Nos. 10475100 and 10505016, the
Knowledge Innovation Project of the Chinese Academy of Sciences
under Grant Nos. KJCX2-SW-N17, KJCX-SYW-N2, Major state basic
research development program under Grant No. 2007CB815000, and the
Helmholtz-DAAD in Germany.

\end{document}